\documentclass[%
 reprint,
nofootinbib,
 amsmath,amssymb,
 aps,
]{revtex4-2}

\usepackage[utf8]{inputenc} % allow utf-8 input
\usepackage[T1]{fontenc}    % use 8-bit T1 fonts
\usepackage{hyperref}       % hyperlinks
\usepackage{url}            % simple URL typesetting
\usepackage{booktabs}       % professional-quality tables
\usepackage{amsfonts}       % blackboard math symbols
\usepackage{nicefrac}       % compact symbols for 1/2, etc.
\usepackage{microtype}      % microtypography
\usepackage{lipsum}
\usepackage{fancyhdr}       % header
\usepackage{blindtext}
\usepackage{physics}
\usepackage{amssymb,amsmath}
\numberwithin{equation}{section}
\usepackage{bbm}
\usepackage{listings}

\usepackage{graphicx}% Include figure files
\usepackage{dcolumn}% Align table columns on decimal point
\usepackage{bm}% bold math
\begin{document}

\preprint{APS/123-QED}

\title{Generalized Parton Distribution Functions via Quantum Simulation of Quantum Field Theory in Light-front Coordinates}% Force line breaks with \\

\author{Carter Gustin}%
 \email{carter.gustin@tufts.edu}
\author{Gary Goldstein}
 \email{gary.goldstein@tufts.edu}

\affiliation{Tufts University}%

\date{\today}% It is always \today, today,

\begin{abstract}
Quantum simulation of quantum field theories offers a new way to investigate properties of the fundamental constituents of matter. We develop quantum simulation algorithms based on the light-front formulation of relativistic field theories. The process of quantizing the system in light-cone coordinates will be explained for a Hamiltonian formulation, which becomes block diagonal, each block approximating the Fock space with a certain harmonic resolution K. We analyze a QCD theory in 2+1D. We compute the analogue of parton distribution functions, the generalized parton distribution functions for mesonic composite particles, like hadrons, in these theories. The dependence of such analyses on the scaling of the number of qubits is compared with other schemes and conventional computations. There is a notable advantage to the light-front formulation.
\end{abstract}

\maketitle

%\tableofcontents
\section{Introduction}
In a 2020 paper by Michael Kreshchuk et. al. \cite{Kreshchuk}, a formalism was laid out to utilize quantum computation to simulate quantum field theory and generate parton distribution functions (PDFs). This paper aims to extend this formalism to higher dimensions and generate generalized parton distributions (GPD) via quantum computing. GPDs describe the momentum distribution of partons in terms of both longitudinal momentum fractions, as well as transverse momentum fractions. We lay out a method to utilize quantum computation to solve for hadronic bound states and use these states to create GPDs. 

A main branch of hadronic physics is to study the internal structure of the nucleon. This was first done via deep inelastic scattering (DIS) ($lp \rightarrow lX$), in which a lepton is scattered off of a nucleon. Measuring the scattered electron's energy and deflection angle gives information about the internal structure of the hadron \cite{DIS}. This process leads to a one dimensional probability distribution, the PDF. The PDF gives the probability to find a quark inside of a hadron with a fraction of the total hadronic momentum. 

The kinematic variables of the GPD are $\{x, \xi, t\}$ where $x$ is the longitudinal momentum fraction of the parton, $\xi$ is the deviation of longitudinal momentum fraction in the process, and $t$ is the total momentum transfer \cite{Mezrag_2022}. The support of the GPD is defined on $(x, \xi) \in [-1, 1]^2$. This leads to three distinct regions: $x \in [\xi, 1]$, $x \in [-\xi, \xi]$, and $x \in [-1, \xi]$. \cite{GPD} The first and third cases are called the DGLAP region which describe the emission and reabsorption of a quark and antiquark respectively while the second case is called the ERBL region, describing the emission of a quark-antiquark pair. In this paper, we are interested in zero-skewness GPDs ($\xi = 0$) for which only the DGLAP region exists. In particular, we show the GPD for a quark inside of a pion, which corresponds to the region $x \in [0, 1]$ only. 

This paper exploits light cone coordinates which describe the motion of a massless particle moving at the speed of light on the light cone. This allows quantum field theory to be formulated similarly to quantum chemistry, which has been simulated on a quantum computer \cite{QuantChem}. Additionally, the vacuum in this formulation is trivial, as the lightfront momentum $P^+$ is bounded below as $P^+ > 0$ strictly. Thus, there is no confusion towards the vacuum state $\ket{0}$ referring to the state with no particles present, and not the state of some hadron with momentum $P = 0$ \cite{vacuum}.

The paper is formatted as follows: In part 2, we lay out the model and formalism used to generate bound states of the quark-antiquark model and solve the corresponding Hamiltonian with VQE. Then in part 3, we describe the GPD in more detail and describe how we can utilize the GPD in fock space. Finally, in part 4, we describe the process for calculating GPDs on a quantum computer and show our result.

\section{Formalism}

\subsection{The Hadronic Model in 2 + 1D}

A fermionic field \cite{QCLC}, $\psi$, interacting with a gluon gauge field, $A_\mu$, can be described by the QCD Lagrangian

\begin{eqnarray}
 \mathcal{L} =  \frac{1}{2} \bar{\psi}(i\gamma^\mu D_\mu - m_q)\psi - \frac{1}{4}G_a^{\mu\nu}G^a_{\mu\nu}
\end{eqnarray}

\noindent where $m_q$ is the bare fermionic mass (for single flavor particles), $G_{\mu \nu}$ is the gluon field strength tensor, $D_\mu = \partial_\mu - igT_aA^a_\mu$ is the covariant derivative, and $g$ describes the coupling between the fields. Light-front (LF) coordinates \cite{Dirac} describe the motion of a (massless) particle moving at the speed of light on the light-cone and will be utilized throughout this paper. Given equal time coordinates $x^\mu = (x^0, x^1, x^2, x^3) = (ct, x, y, z)$, the coordinate transformation to LF coordinates is: $x^{\pm} = x^0 \pm x^3.$
    
The fields \cite{QCLC} are given as

    \begin{equation}
        A_\mu(x) = \sum_{n = 1}^\Lambda \frac{1}{(\Omega n)^{\frac{1}{2}}} (a_n\epsilon_\mu(k, \lambda)e^{-ik_n^\nu x_\nu} + a_n^\dagger\epsilon^*_\mu(k, \lambda) e^{ik_n^\nu x_\nu})
    \end{equation}

\noindent and 
    \begin{equation}
        \psi(x) = \sum_{n = 1}^\Lambda \frac{1}{(\Omega)^{\frac{1}{2}}} (u(k)b_ne^{-ik_n^\nu x_\nu} + v(k)d_n^\dagger e^{ik_n^\nu x_\nu})
        \label{fields}
    \end{equation}

\noindent with discretized single-particle light-cone momenta $k_n^+ = \frac{2\pi}{L}n;$ $n = 1, 2, ...\Lambda$ where $\Lambda$ is the momentum cut-off which must be chosen by hand, and $L$ is the size of the box which the fields are quantized into. Luckily, in LF coordinates, there are three constants of motion (which constitute a complete set of commuting observables): $Q$ the charge, $P^+$ the LF momentum, and $P^-$ the LF energy, which isolate the dependence of the box size, $L$. In LF coordinates, $Q = \sum_n (b_n^\dagger b_n - d_n^\dagger d_n)$, $ P^- = \frac{L}{2\pi}H$,$P^+ = \frac{2\pi}{L}K$, where a new operator, $K$ the harmonic resolution $K = \sum_n n(a_n^\dagger a_n + b_n^\dagger b_n + d_n^\dagger d_n)$  gives the total momentum of the hadron. Physically, one can think of each parton inside the hadron having light-cone momentum $k^+$ and the harmonic resolution $K$ (the total LF hadronic momentum) is the sum of all of the individual parton momenta. 
 
 \subsection{Fock States}
 We look at Fock states in second-quantized momentum space for fermions, antifermions and bosons \cite{Kreshchuk}: 

    \begin{equation}
        \ket{i} = \ket*{n_1, \dots n_N; 
            \Bar{n}_1, \dots, \Bar{n}_{\Bar{N}};
            \Tilde{n}_1^{\Tilde{m}_1}, \dots,
            \Tilde{n}_{\Tilde{N}}^{\Tilde{m}_{\Tilde{N}}}}
    \end{equation}

\noindent which can be built up from the vacuum state $\ket{0}$ via creation opperators $a_i^\dagger, b_i^\dagger, d_i^\dagger$ subject to the constraint $\sum_i^{N + \Bar{N} + \Tilde{N}} \hat{n}_i^+ = K$ and $\sum_i k_{i, \perp} = P_\perp$ where $P_\perp$ is the total transverse momentum. The first $N$ terms refer to fermions with $n_i \in \{0, 1\}$, the second $\Bar{N}$ terms refer to antifermions again with $\Bar{n}_i \in \{0, 1\}$, while the final $\Tilde{N}$ terms refer to bosons with $\Tilde{m}_i$ bosons in a given mode. There is no explicit cutoff on bosonic occupancy manually inputted; however, $K$ serves as a cutoff as there cannot be more than $K$ total particles. Note that only occupied modes are shown in a given Fock state. The set of quantum numbers that each fermion can take on are $\{n^+, \boldsymbol{n_\perp}, \lambda, c, f \}$ where $\lambda$ refers to the helicity of the fermion, $c$ is the color, and $f$ is the flavor. In a given Fock state, instead of each fermionic mode representing a single quantum number, each fermionic mode represents a set of numbers.

This can obviously increase the number of qubits quite rapidly as there are 2 possible helicities, 3 possible colors and 6 possible flavors for each $n^+, \boldsymbol{n}_\perp$ so this grows quickly as $n^+, \boldsymbol{n}_\perp$ increase.
There are a few ways to decrease the number of qubits that must be used to represent this system. First, we use the Tamm-Dancoff cutoff which looks only at the $\ket{q \Bar{q}}$ or  $\ket{q \Bar{q} g}$ sectors (for the purposes of this paper, we only look at the $\ket{q \Bar{q}}$ sector). With this cutoff, we are essentially looking at mesonic bound states such as the pion. Additionally, we disregard the helicity, color and flavor and only keep the quantum numbers $n^+$ and $\boldsymbol{n}_\perp$. In order to reduce resources, we utilize a binary encoding, where each of the $N$ possible fock states is written in binary form.

\subsection{The Bound State problem}
In this section, the bound state problem will be discussed (note that much of this information comes from the Pauli, Brodsky, Pinsky paper \cite{QCLC} and the W{\"o}lz paper \cite{Wolz}). The eigenvalue problem we must solve is 

\begin{equation}
    H\ket{\Psi} = \frac{M^2 + P^2_\perp}{2P^+}\ket{\Psi}.
\end{equation}

$\ket{\Psi}$ is the hadronic statefunction defined in the set of basis Fock states $\ket{\mu_n}$ for the partons. A given Fock state has the form: 

\begin{equation}
    \ket*{q \Bar{q}: k_i^+, \vec{k}_{i,\perp}} = b^\dagger(q_1)d^\dagger(q_2)\ket{0}
\end{equation}

\noindent where $b^\dagger(q_1)$ creates a quark with the quantum numbers $q_1$ and $d^\dagger(q_2)$ creates an antiquark with the quantum numbers $q_2$.
There are a finite number of Fock states of this form satisfying $\sum_i k^+_i = P^+$ and $\sum_i k_{i,\perp} = P_\perp$. This set of Fock states forms a basis to which the hadronic state $\Psi$ can be written:

$$\ket{\Psi} = \sum_n \bra{\mu_n}\ket{\Psi}\ket{\mu_n}. $$

Solving the bound state problem using VQE will allow us to determine the ground state energy as well as the coefficients $\bra{\mu_n}\ket{\Psi}$ in the expansion of $\ket{\Psi}$ which allows us to write down the ground state wavefunction. 

\subsection{Solving the Hamiltonian with VQE}

One commonly used algorithm to find the ground state of a Hamiltonian is the Variational Quantum Eigensolver (VQE) \cite{VQE}\cite{QuantChem}. This algorithm relies on the variational principle of quantum mechanics which states that the expectation value of the Hamiltonian for a given state, $\ket{\psi}$, must be greater than or equal to the ground state energy: 
$$ \matrixel{\psi}{H}{\psi} \geq E_0.$$

We can generate a parameterized state on a quantum computer, $\ket*{\psi\vec{(\theta)}}$, such that as we alter the values of $\vec{\theta} = (\theta_1, \theta_2, ..., \theta_n)$, we expect to obtain a better estimate on the ground state energy of $H$. Known as a hybrid algorithm for utilizing both a classical and quantum processor to solve for the ground state, the steps are the following: 

    \begin{enumerate}
    
        \item Define a state/operator mapping from the physical states in second-quantized space to qubit states. 
        \item Choose an initial ansatz state that can be easily prepared with a quantum computer $\ket{\psi_0}$ (Here this is taken as a fock state in the (K, Q)-sector of the Hamiltonian). 
        \item Create a parameterized circuit $U(\vec{\theta})$ that prepares the parameterized state via $\ket*{\Psi(\vec{\theta})} = U(\vec{\theta})\ket{\psi_0}$\footnote{In a direct "one-hot" encoding of Fock states to qubit states, we use a UCC ansatz. Here in a binary encoding, a hardware efficient ansatz is used. See \ref{fig:circuit.png}}. 
        \item Calculate $E(\vec{\theta}) = \matrixel*{\psi(\vec{\theta})}{H}{\psi(\vec{\theta})}$ on a quantum processor for a given set of parameters $\vec{\theta}$. 
        \item Use a classical processor to update $\vec{\theta}$ based on the current value of $E(\vec{\theta})$.
        \item Repeat this process until convergence (within a set tolerance) to the ground state $E_0$
    \end{enumerate}

\subsection{2 + 1D Hamiltonian}

The 2 + 1D Yukawa Hamiltonian is given by
\begin{equation}
    H_{LC} = P^\nu P_\nu = P^+P^- - P^2_\perp
\end{equation}
where $P^+ (= K)$ and $P_\perp$ are chosen parameters. 
%$P^+ = K$ and 
%It is permissible to take $P_\perp = 0$ as 
There exists a reference frame such that the total hadronic transverse momentum is zero, but the individual partonic transverse momenta are non-zero. The Hamiltonian $H_{LC} = KP^-$ 
%where $P^-$ is more complicated relative to the 1 + 1D Hamiltonian.
is given explicitly in Pauli, Brodsky, Pinsky\cite{QCLC}.
It is the sum of a kinetic term and an interaction (potential) term $H = T + U$ with the kinetic piece diagonal in number operators: 
\begin{equation}
    T = \sum_q \frac{m^2_q + k_\perp^2}{x}\left(a^\dagger_q a_q + b^\dagger_q b_q+ d^\dagger_q d_q\right)
\end{equation}
where $x = n^+/K$ and the sum goes over the set of all possible quantum numbers, $q = n^+, {\bf n}_\perp, \ldots$ 
%\begin{equation*}
 %   \sum_q = \sum_{n^+}\sum_{\boldsymbol{n}^\perp}\dots
%\end{equation*}
where the dots include any other quantum numbers not used in our problem (such as helicity, color, flavor). 
The interaction term is a sum of four pieces: $U = S + C + V + F$. See the Pauli, Brodsky, Pinsky \cite{QCLC} paper for the full interaction matrix for an SU(N)-Meson. In the $\ket{q \Bar{q}}$ sector we are interested in, the only part of the interaction that comes into play is the seagull part (S) given as: 

%\begin{widetext}
%
%\begin{equation}
 %   S_{q\Bar{q}} = \sum_{q_1^\prime, q_2^\prime, q_1, q_2} \frac{2g^2P^+\Delta}{2\Omega} \left( \frac{1}{(x_1 + x_2)^2 - (m_g/\Omega)^2} - \frac{1}{(x_1 - x_3)^2 - (m_g/\Omega)^2} \right )b_{1^\prime}^\dagger d_{2^\prime}^\dagger b_{1} d_{2} 
%\end{equation}
%\end{widetext}

\begin{equation}
    S_{q\Bar{q}} = \sum_{q_1^\prime, q_2^\prime, q_1, q_2} \frac{2g^2P^+\Delta}{2\Omega} \left( \frac{1}{(x_1 - x_3)^2 - (m_g/\Omega)^2}\right)b_{1^\prime}^\dagger d_{2^\prime}^\dagger b_{1} d_{2} 
\end{equation}

\noindent where
$\Delta = \delta(k^+_{1^\prime} + k^+_{2^\prime} |k^+_{1} + k^+_{2})\delta^2(k^\perp_{1^\prime} + k^\perp_{2^\prime} |k^\perp_{1} + k^\perp_{2}).$ and $\Omega$ is the normalization volume depending on the box lengths that we define our fields in. $b_{1} d_{2}$ annihilates a quark and an antiquark with the quantum numbers described by $q_1, q_2$ while $b_{1^\prime}^\dagger d_{2^\prime}^\dagger$ creates a quark and an antiquark with the quantum numbers described by $q_{1^\prime}, q_{2^\prime}$. The additional $(m_g/\Omega)^2$ term is added to the exchange gluon propagator to remove the singularity (this is done by assuming a vanishingly small gluonic mass).  With the explicit form of the Hamiltonian in the $\ket{q \Bar{q}}$ sector, we can write the Hamiltonian in the basis of Fock states\cite{matel}: $\matrixel{\mu_m}{H}{\mu_n}.$
From here, VQE is utilized to find the ground bound state of H which is then used to find the GPD. 

\begin{figure*}[t]
\includegraphics[height = 3cm, width = 18cm]{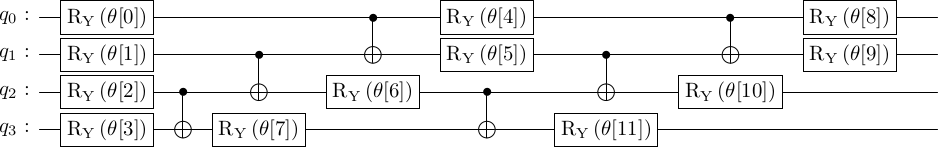}
    \caption{VQE Circuit for $\ket{q\bar{q}}$ with $P^+ = 4, P^\perp = 0, \Lambda_\perp = 2$ via \texttt{qiskit.RealAmpltidues} with 2 repetition layers}
    \label{fig:circuit.png}
\end{figure*}

\noindent Each fock state contains a quark and an antiquark, each with longitudinal momentum $k_i$ and transverse momentum $k_i^\perp$. Given hadronic momenta $P^+, P^\perp$, there are a finite number of fock states. From here, a binary encoding to qubit states is utilized in order to reduce qubit count. Figure \ref{fig:circuit.png} shows the ansatz used for VQE (\texttt{qiskit.circuit.library.RealAmpltidues}). The parameters of the model are $P^+ = 4, P^\perp = 0, \Lambda_\perp = 2$, we get 15 fock states which can be mapped to 4 qubits ($4 = \lceil \log_2(15) \rceil$).

% \begin{widetext}
% \end{widetext}

\section{The Generalized Parton Distribution Function (GPD)}
\begin{figure}[h]
    \centering
    \includegraphics[width = \columnwidth]{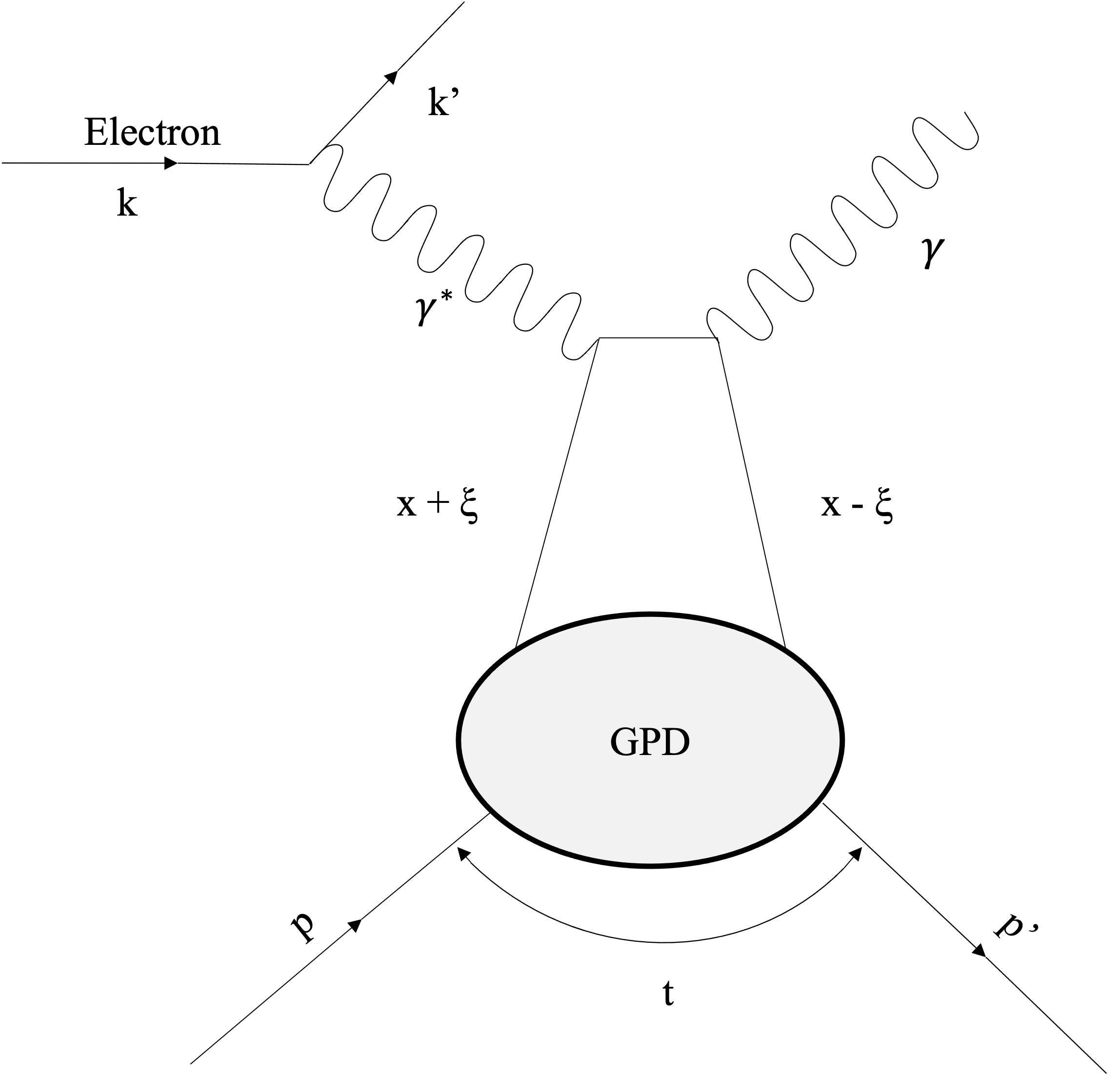}
    \caption{The Handbag Diagram for Deeply Virtual Compton Scattering.}
    \label{fig:Handbag}
\end{figure}

The GPD \cite{GPD} is an extension of the canonical parton distribution function (PDF). The PDF represents the probability of finding a parton inside of a hadron with a fraction of the total hadronic momentum. The Deeply Virtual Compton Scattering (DVCS) \ref{fig:Handbag} process allows for a much more comprehensive understanding of the hadron. In DVCS, the hadron's internal structure is determined by probing via a high-energy electron beam. Probing the hadron can be quite challenging due to continuous interactions between quarks and gluons as well as particle creation/annihilation; however, asymptotic freedom in QCD can be exploited \cite{Mezrag_2022}. Asymptotic freedom is a property of some gauge theories (in this case QCD) that leads to a decrease in coupling between quarks and gluons to decrease at high energies. In this regime, the interaction can be studied perturbatively. 

In the handbag diagram of figure \ref{fig:Handbag}, a lepton emits a highly virtual photon such that the momentum transfer scale is $Q^2 = -q^2 >> M_N^2$, where $q^2$ is the momentum of the virtual photon.The virtual photon is now able to probe one individual parton which receives a large amount of momentum transferred to it. For the parton to remain inside of the hadron and not hadronise, the parton must release energy, which it does through the emission of a real photon with momentum $q'$. This real photon carries the information on the quark's transverse and longitudinal momentum \cite{Nature}. The Lorentz invariant parameters defined in the GPD, in addition to the standard momentum fraction $x$ of the PDF are $\xi$ and $t$ defined as

\begin{equation}
\xi = -\frac{\Delta^+}{2P^+}, \quad t = \Delta^2.
\end{equation}

$\xi$ describes to the change in longitudinal momentum after the quark is reabsorbed in the hadron while $t$ is the invariant momentum transfer. 
In what follows, we lay a methodology to calculate GPDs using quantum computation. We exploit the hermiticity and unitarity of the Pauli operators and the \texttt{SWAP test} to calculate off-diagonal matrix elements.

\subsection{Formal Definition of the GPD}
The quark distributions are defined by M. Diehl \cite{GPD} as 

\begin{equation}
    H^q = \frac{1}{2}\int \frac{dz^-}{2\pi}e^{ixP^+z^-}\matrixel{P^\prime}{\Bar{q}(-\frac{z}{2})\gamma^+q(\frac{z}{2})}{P}\vert_{z^+ = \boldsymbol{z} = 0}
    \label{GPD}
\end{equation}

\noindent where $q(\frac{z}{2})$ is a quark field operator and $\gamma^+$ is one of the LF Dirac matrices defined by $\gamma^+ = \gamma^0 + \gamma^3.$ In order to put equation \ref{GPD} to use, we must write the quark field operators in terms of fermionic and bosonic creation and annihilation operators as well as define the Fock states in 2 + 1D formalism. Note that in the limit that $t = 0$, (i.e. the incoming and outgoing hadronic states are the same: $\ket{P} = \ket{P'}$), we 
obtain the formal definition of the PDF and hence: $H^q(x, 0, 0) = f_q(x)$
\noindent where $f_q(x)$ is the parton distribution function 
\begin{equation}
    f_q(x) = \matrixel{\Psi_{K,Q}}{N_q}{\Psi_{K,Q}}.
    \label{PDF}
\end{equation}

\noindent Here, $N_q$ refers to a number operator for a $q$-type particle (e.g. fermion, antifermion, boson). The PDF is a probability distribution so it is normalized by particle number; however, the GPD is not a squared ampltitude, thus it is not considered a probability distribution. 

\subsection{GPD in terms of Fock operators}

Golec-Biernat and Martin (as well as Ji)\cite{OffDiag} \cite{Ji} give the explicit form of the GPD (off-diagonal distribution) in terms of creation/annihilation operators by writing the quark field operators $q(\frac{-z}{2})$ and $q(\frac{z}{2})$ (defined in equation \ref{fields}  )in terms of fermionic and bosonic operators. Eq \ref{GPDFock} depends on the DGLAP and ERBL regions and we are interested in the quark GPD so we look at the region $\theta(x \geq \xi)$.
\begin{widetext}
\begin{eqnarray}
    \label{GPDFock}
    H^q(x, \xi) &&= \frac{1}{2\Bar{P}^+}\int \frac{d^2k_T}{2\sqrt{\abs{x^2-\xi^2}(2\pi)^3}} \\ \nonumber
    && \sum_{\lambda} [\bra{P^\prime}b_\lambda^\dagger((x - \xi)\Bar{P}^+, k_T - \Delta_T)b_\lambda((x + \xi)\Bar{P}^+, k_T)\ket{P} \theta(x \geq \xi) \\ \nonumber
    && + \bra{P^\prime}d_\lambda^\dagger((-x + \xi)\Bar{P}^+, -k_T + \Delta_T)b_{-\lambda}((x + \xi)\Bar{P}^+, k_T)\ket{P} \theta(-\xi < x < \xi) \\ \nonumber
    && - \bra{P^\prime}d_\lambda^\dagger((-x - \xi)\Bar{P}^+, k_T - \Delta_T)d_{\lambda}((-x + \xi)\Bar{P}^+, k_T)\ket{P} \theta(x \leq \xi)].
\end{eqnarray}
\end{widetext}

This is called the off-diagonal distribution because it connects different Fock states $\ket{P}$ and $\ket{P^\prime}$ whereas the PDF (forward distributions) is diagonal and connects the same states. $H^q$ simplifies when we look at the zero skewness case ($\xi = 0)$ since we are interested in hadrons with the same resolutions before and after an interaction.

$\ket{P}$ is the hadronic bound state that is obtained by running VQE with the 2 + 1D Hamiltonian above. This state will be a superposition of Fock states defined for a particular $\ket{q \Bar{q}}$ sector depending on $(P^+, P_\perp)$. $\ket{P^\prime}$ can be any chosen output state (not all output states will give non-zero GPDs as not all states can be "reached" by the interaction from a photon). The four-vector $P^{\prime \mu} = (P^{\prime +},P^{\prime -}, P^{\prime}_{\perp})$ allows us to choose $P^{\prime +}$ and $P^{\prime}_{\perp}$ and subsequently create a new set of Fock states $\{ \ket{\mu_n^\prime} \}$ that satisfy the parameters given by $P^{\prime \mu}$. $\ket{P'}$ is a sum of all kets that satisfy the constraints $\sum_i^{N + \Bar{N} + \Tilde{N}} \hat{n}_i^+ = K$ and $\sum_i k_{i, \perp} = P_\perp$, normalized to unity.

\section{Calculating the GPD on a Quantum Computer}

The main goal of this paper is to show a methodology of obtaining GPDs on a quantum computer. Solving for the bound states is done by solving the lightfront Schr\"{o}dinger equation $H_{eff}\ket{\Psi} = M^2 \ket{\Psi}$, whose ground state $\ket{\Psi_0}$ can be found on a quantum computer via VQE. The ground state of the hadron found with VQE will be a state of the form $\ket{P}$ which can be used to calculate hadronic distributions (PDF, GPD, FF). The parton distribution function is diagonal in that it connects states with same longitudinal and transverse momentum (see equation \ref{PDF}). Thus, it can be computed on a quantum computer using the Hadamard test \cite{Hadamard} which can approximate $\operatorname{Re\expval{U}{\psi}}$. GPDs, however, are commonly referred to as off-forward parton distributions because they connect states with different momentum: $\ket{P}$ and $\ket{P'}$. There is not an equivalent test for matrix elements off the diagonal, so in order to calculate the GPD $\bra{P'}U\ket{P}$ on a quantum computer, we must exploit the \texttt{SWAP test} \cite{SWAP}.

\begin{figure}[h]
    \centering
    \includegraphics[width = 7cm, height = 3cm]{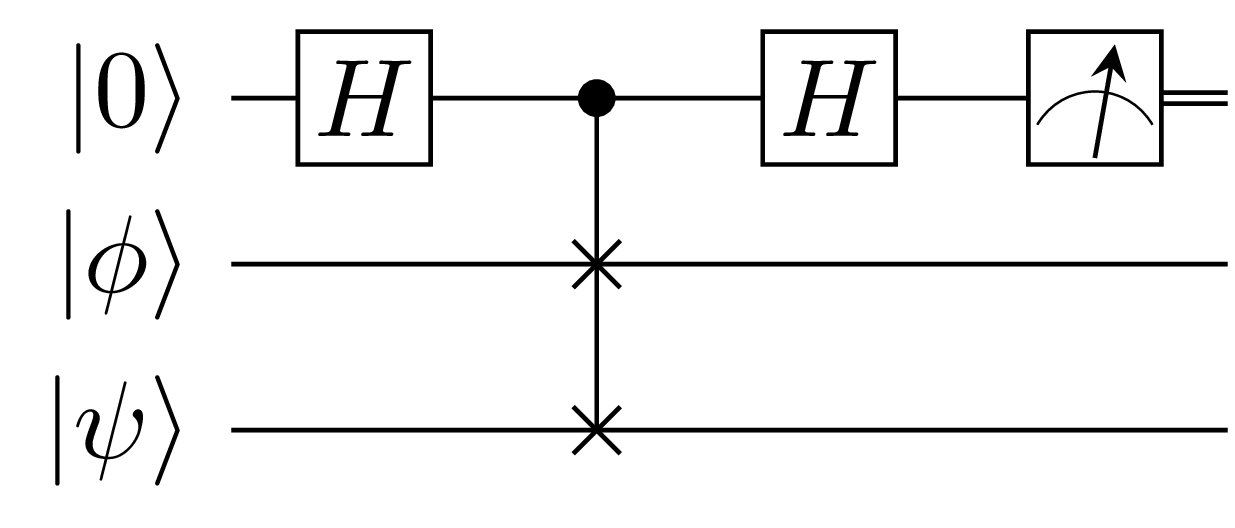}
    \caption{The SWAP Test \cite{swapimg}}
    \label{fig:swaptest}
\end{figure}

The \texttt{SWAP test} uses an ancilla qubit to measure the overlap between two quantum states $\abs{\bra{\phi}\ket{\psi}}^2$ without destroying the information of the states. Figure \ref{fig:swaptest} shows how this is accomplished. 

We can initialize the circuit to have $\ket{\phi} = \ket{\Bar{P}}$ and $\ket{\psi} = \ket{P'}$ using \texttt{qiskit.initialize} which uses the methods outlined in \cite{initialize}. Note that $\ket{\Bar{P}}$ is the state such that $\ket{\Bar{P}} = U\ket{P}$ where $U$ is the operator defined in \ref{GPDFock} $U \equiv b_\lambda^\dagger((x - \xi)\Bar{P}^+, k_T - \Delta_T)b_\lambda((x + \xi)\Bar{P}^+, k_T)$. $U \in P^{\otimes m}$ (the Pauli group on $m$ qubits) so it is Hermitian and unitary. 

After running VQE with the circuit in figure \ref{fig:circuit.png} and preparing the states $\ket{\bar{P}}$ and $\ket{P'}$ with an ancilla qubit $\ket{0}$, we can use the \texttt{SWAP test} to measure $\abs{\bra{P'}\ket{\bar{P}}}^2$ whose square root will give us an approximation to the GPD for particular momentum fractions $\{x, t\}$. In doing this, the following figure \ref{fig:gpd} shows the GPD for a quark. 

\begin{figure}[h]
    \includegraphics[height = 8cm, width = 10cm]{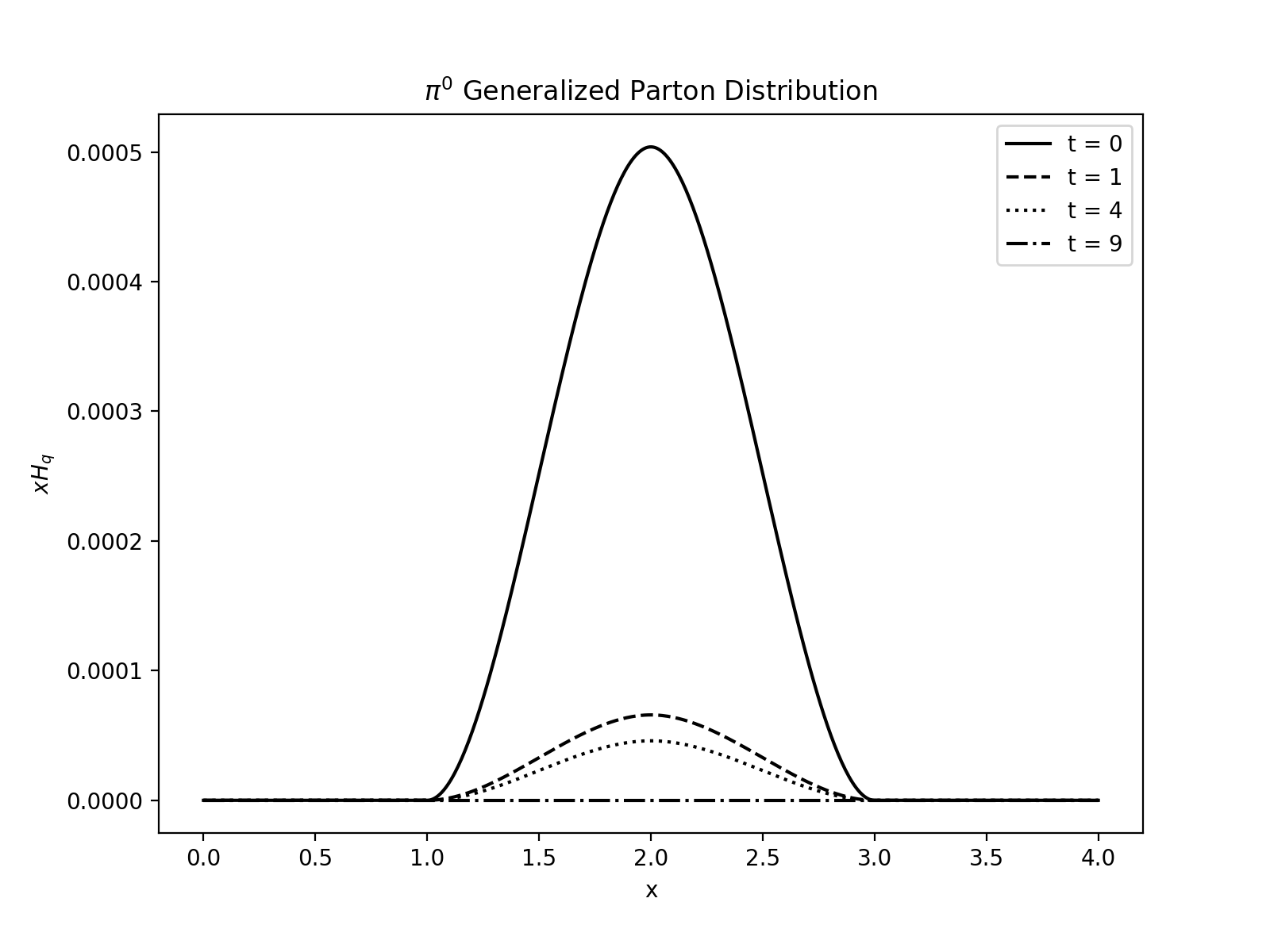}
    \caption{$\pi^0$ quark GPD}
    \label{fig:gpd}
\end{figure}

The GPD collapses to the PDF at $t = 0$ and falls off with increasing $t$. This corresponds to the fact that the quark in the pion has largest probability to be found with no transverse momentum. This is consistent with the condition that $t$ be small \cite{GPD}. The antiquark GPD will be the same as the quark GPD because there are no gluons in this system.  

\section{Summary}

The GPD is a transition amplitude of transitioning from the state $\ket{P}$ to the state $\ket{P^\prime}$ defined above. In the zero-skewness case $\xi = 0$, the only states that the hadron can transition to are states with the same light-front momentum but with different transverse momentum. 

By concentrating on the $q \Bar{q}$  sector, we have simulated a $\pi^0$ type hadron, whose wave function is predominantly valence quarks. While we have not solved the full QCD Hamiltonian, using light front quantization, we have shown how a simpler quantum field theory gives rise to hadronic states composed of Fock states. Those states are mapped onto qubit states - the state preparation being a crucial part of the connection to quantum computers. From the VQE optimization, states are constructed whose resulting PDFs and GPDs can be extracted, ultimately by quantum computers. The resulting distributions show very reasonable resemblances to known measured distribution functions. 
In future work, by extending the Hamiltonian to include flavors and colors, as well as a third space dimension, a more realistic representation of hadrons via quantum computers will be sought. Additionally, we hope to study baryonic and quark-diquark model of bound states.

\begin{acknowledgements}
We have benefited from communications with our quantum computer group colleagues at Tufts and with S. Brodsky, G. deTeramond, J. Vary.
\end{acknowledgements}

\appendix

\bibliography{GPD}

\end{document}